\documentclass{article}

\title{Courcelle's theorem for triangulations}
\author{Benjamin A. Burton\thanks{School of Mathematics and Physics, The University of Queensland,
Brisbane, QLD 4072, Australia, \texttt{bab@maths.uq.edu.au}.
Supported by the Australian Research Council (projects DP1094516, DP110101104).
A full version of this paper is available at \texttt{arXiv:1403.2926}.}}

\begin{document}

\maketitle

\begin{abstract}
In graph theory, Courcelle's theorem essentially states that, if an
algorithmic problem can be formulated in monadic second-order logic,
then it can be solved in linear time for graphs of bounded treewidth. We
prove such a metatheorem for a general class of triangulations of
arbitrary fixed dimension $d$, including all triangulated $d$-manifolds: if
an algorithmic problem can be expressed in monadic second-order logic,
then it can be solved in linear time for triangulations whose dual
graphs have bounded treewidth.
%We give several applications to 3-manifold topology, a setting with many
%difficult computational problems but very few known parameterised complexity
%results.
%
%We apply our results to 3-manifold topology, a setting with many
%difficult computational problems but very few parameterised complexity
%results, and where treewidth has practical relevance as a parameter.
%Using our metatheorem, we recover and generalise earlier
%fixed-parameter tractability results on taut angle structures and
%discrete Morse theory respectively, and prove a new fixed-parameter
%tractability result for computing the powerful but complex Turaev-Viro
%invariants on 3-manifolds.
%
This is joint work with Rodney G.~Downey.
\end{abstract}

%\usepackage{amsmath}
%\usepackage{amssymb}
%\usepackage{graphicx}

%%%%%%%%%%%%%%%%%%%%%%%%%%%%%%%%%%%%%%%%%%%%%%%%%%%%%%%%%%%%%%%%%%%%%%%%
%
%   Theorems and lemmata
%
%%%%%%%%%%%%%%%%%%%%%%%%%%%%%%%%%%%%%%%%%%%%%%%%%%%%%%%%%%%%%%%%%%%%%%%%

% Custom theorem types:
%
% \newtheorem{corollary}[theorem]{Corollary}
% \newtheorem{lemma}[theorem]{Lemma}
%\spnewtheorem{algorithm}[theorem]{Algorithm}{\bfseries}{\itshape}
%\spnewtheorem{babproblem}[theorem]{Problem}{\bfseries}{\itshape}

% \theoremstyle{definition}
% \newtheorem{defn}[theorem]{Definition}
% \newtheorem{example}[theorem]{Example}
%\spnewtheorem*{notation}{Notation}{\bfseries}{\upshape}

%%%%%%%%%%%%%%%%%%%%%%%%%%%%%%%%%%%%%%%%%%%%%%%%%%%%%%%%%%%%%%%%%%%%%%%%
%
%   Custom macros and environments
%
%%%%%%%%%%%%%%%%%%%%%%%%%%%%%%%%%%%%%%%%%%%%%%%%%%%%%%%%%%%%%%%%%%%%%%%%

% Custom macros for general use within the text:
%
\newcommand{\adj}{\mathop{adj}}
\newcommand{\col}{\mathop{col}}
\newcommand{\co}{\colon\thinspace}
\newcommand{\dual}{\mathcal{D}}
\newcommand{\hasse}{\mathcal{H}}
\newcommand{\of}{\!:\!}
\newcommand{\inc}{\mathop{inc}}
\newcommand{\isarc}{\mathop{is\_arc}}
\newcommand{\iscol}{\mathop{is\_col}}
\newcommand{\isface}{\mathop{is\_face}}
\newcommand{\isnode}{\mathop{is\_node}}
\newcommand{\mfd}{\mathcal{M}}
\newcommand{\ms}{\mathop{MS}}
\newcommand{\tri}{\mathcal{T}}
\newcommand{\tw}{\mathrm{tw}}

%%%%%%%%%%%%%%%%%%%%%%%%%%%%%%%%%%%%%%%%%%%%%%%%%%%%%%%%%%%%%%%%%%%%%%%%
%
%   Introduction
%
%%%%%%%%%%%%%%%%%%%%%%%%%%%%%%%%%%%%%%%%%%%%%%%%%%%%%%%%%%%%%%%%%%%%%%%%

\section{Introduction}

Parameterised complexity is a relatively new and highly successful
framework for understanding the computational complexity of hard problems
for which we do not have a polynomial-time algorithm.
The key idea is to measure the complexity not just in terms of the input
size (the traditional approach), but also in terms of
additional \emph{parameters} of the input or of the problem itself.
As a result, even if a problem is (for instance) NP-hard,
we gain a richer theoretical understanding of those classes of inputs for
which the problem is still tractable, and we acquire new practical tools
for solving the problem in real software.

%For example,
%finding a Hamiltonian cycle in an arbitrary graph is NP-com\-plete,
%but for graphs of fixed treewidth $\leq k$
%it can be solved in linear time in the input size
%\cite{downey99-param}.
A problem is called \emph{fixed-parameter tractable} in the
parameter $k$ if, for any class of inputs where $k$ is universally
bounded, the running time becomes polynomial in the input size.
Treewidth in particular
% (which roughly measures how ``tree-like'' a graph is
% \cite{robertson86-algorithmic})
is extremely useful as a parameter.
% Roughly speaking, the treewidth of a graph
% \cite{robertson86-algorithmic}
% measures how ``tree-like'' the graph is: all
% trees have treewidth $1$, and the complete graph $K_n$
% has treewidth $n-1$.
A great many graph problems are known to
be fixed-parameter tractable in the treewidth,
in a large part due to Courcelle's celebrated ``metatheorem''
\cite{courcelle87-context-free,courcelle90-rewriting}:
for \emph{any} decision problem $P$ on graphs,
if $P$ can be framed using monadic second-order logic,
then $P$ can be solved in \emph{linear time} for graphs of
universally bounded treewidth $\leq k$.

Our motivation here is to develop the tools of
parameterised complexity for systematic use in the field of
geometric topology, and in particular for 3-manifold topology.
%This is a field with natural and fundamental algorithmic problems,
%such as determining whether two knots or triangulations
%are topologically equivalent,
%and in three dimensions such problems are often % decidable but
%extremely complex \cite{matveev03-algms}.
%
Here parameterised complexity is appealing as a theoretical
framework for identifying when ``hard'' topological problems can be
solved quickly.  Unlike average-case complexity or
generic complexity, it avoids the need to work with
\emph{random inputs}---something that still poses major difficulties
for 3-manifold topology.
The viability of this framework is shown by recent
parameterised complexity results in topological settings such as
knot polynomials \cite{makowsky05-tutte,makowsky03-knot},
angle structures \cite{burton13-taut},
discrete Morse theory \cite{burton13-morse}, and
3-manifold enumeration \cite{burton14-fpt-enum}.

The treewidth parameter plays a key role in all of the results above.
For topological problems whose input is a triangulation $\tri$,
we measure the treewidth of the \emph{dual graph} $\dual(\tri)$,
whose nodes describe top-dimensional simplices of
$\tri$, and whose arcs show how these simplices are joined together
along their facets.
In 3-manifold topology this parameter has a natural interpretation,
and there are common settings in which the treewidth remains small.

Our main result is a Courcelle-like metatheorem
for use with triangulations.  Specifically,
we describe a form of monadic second-order
logic on triangulations of fixed dimension $d$, and
show that all problems expressible in this
logical framework are fixed-parameter tractable in the treewidth of the
dual graph of the input triangulation.
We apply this
to discrete Morse theory in arbitrary dimensions,
and to computing the powerful
Turaev-Viro invariants of 3-manifolds.

% We recover earlier results on
% taut angle structures \cite{burton13-taut} and discrete Morse theory
% \cite{burton13-morse}, generalise the latter result

% We emphasise that our results
% depend crucially on how we define a triangulation.
% We cannot use arbitrary simplicial complexes, where low-dimensional
% faces can be joined together independently of
% the larger simplices to which they belong.
% Instead our triangulations are formed purely by joining together
% $d$-simplices along their $(d-1)$-dimensional facets.
% This definition is flexible enough to encompass any reasonable
% concept of a triangulated $d$-manifold.  Indeed, it covers
% structures beyond simplicial complexes,
% such as the highly efficient
% \emph{one-vertex triangulations} % \cite{jaco03-0-efficiency}
% and \emph{ideal triangulations} % \cite{thurston78-lectures}
% favoured by many 3-manifold topologists.

%%%%%%%%%%%%%%%%%%%%%%%%%%%%%%%%%%%%%%%%%%%%%%%%%%%%%%%%%%%%%%%%%%%%%%%%
%
%   Preliminaries
%
%%%%%%%%%%%%%%%%%%%%%%%%%%%%%%%%%%%%%%%%%%%%%%%%%%%%%%%%%%%%%%%%%%%%%%%%

\section{Triangulations}

We first describe the general class of $d$-dimensional triangulations
with which we work.
In essence, these triangulations are formed by identifying (or ``gluing'')
facets of $d$-simplices in pairs.
This definition does not cover all simplicial complexes (in which
lower-dimensional faces can also be identified independently), but it does
encompass any reasonable definition of a triangulated $d$-manifold;
moreover, it allows more general structures that simplicial complexes
do not, such as the highly efficient
``1-vertex triangulations'' and ``ideal triangulations''
often found in algorithmic 3-manifold topology.
% \cite{jaco03-0-efficiency,thurston78-lectures}.
The details follow.

Let $d \in \N$.  A \emph{$d$-dimensional triangulation}
consists of a collection of abstract $d$-simplices $\Delta_1,\ldots,\Delta_n$,
some or all of whose facets (i.e., $(d-1)$-faces)
are affinely identified in pairs.
Each facet $F$ of a $d$-simplex may only be identified with at most one
other facet $F'$ of a $d$-simplex; this may be another facet
of the \emph{same} $d$-simplex, but it cannot be $F$ itself.
%Those facets that are not identified with any other facet together form
%the \emph{boundary} of the triangulation.

Consider any integer $i$ with $0 \leq i < d$.
There are $\binom{d+1}{i+1}$ distinct $i$-faces of each simplex
$\Delta_1,\ldots,\Delta_n$.
As a consequence of the facet identifications, some of these $i$-faces
become identified with each other; we refer to each class of
identified $i$-faces as a single \emph{$i$-face of the triangulation}.
As usual, 0-faces and 1-faces are called \emph{vertices} and
\emph{edges} respectively.
A \emph{simplex of the triangulation} explicitly refers to one of the
$d$-simplices $\Delta_1,\ldots,\Delta_n$ (not a smaller-dimensional face),
and for convenience we also refer to these as
\emph{$d$-faces of the triangulation}.

A \emph{$d$-manifold triangulation} is simply a $d$-dimensional triangulation
whose underlying topological space is a $d$-manifold when using the
quotient topology.

By convention, we label the vertices of each simplex as $0,\ldots,d$.
We also arbitrarily label the vertices of each $i$-face of the
triangulation as $0,\ldots,i$ (e.g., for $i=1$ this corresponds
to placing an arbitrary direction on each edge).
% Note that there are many possible ways in which the
% vertex labels of an $i$-face of the triangulation
% might correspond to vertex labels on the constituent simplices.

\begin{figure}[tb]
    \centering
    \subfigure[A Klein bottle $\mathcal{K}$\label{fig-kb}]{%
        ~~~\includegraphics[scale=0.7]{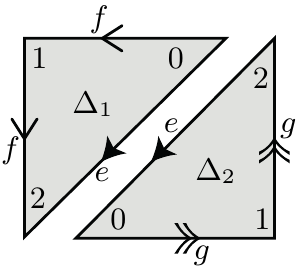}~~~}
    \hspace{2cm}
    \subfigure[The dual graph $\dual(\mathcal{K})$\label{fig-dual}]{%
        \qquad\includegraphics[scale=0.7]{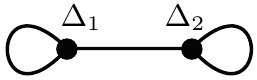}\qquad}
    \caption{A 2-dimensional triangulation}
\end{figure}

% \begin{example} \label{ex-tri}
    Figure~\ref{fig-kb} illustrates a 2-manifold triangulation with
    $n=2$ simplices whose
    underlying topological space is a Klein bottle.  As indicated by the
    arrowheads, we identify the following pairs of facets (i.e., edges):
    $\Delta_1\of02 \longleftrightarrow \Delta_2\of20$;
    $\Delta_1\of01 \longleftrightarrow \Delta_1\of12$; and
    $\Delta_2\of01 \longleftrightarrow \Delta_2\of12$.
    The resulting triangulation has one vertex (since
    all vertices of $\Delta_1$ and $\Delta_2$ become identified together),
    and three edges (labelled $e,f,g$ in the diagram).

    %If we label each edge so that vertices $0$ and $1$ are at the
    %base and the tip of the arrow respectively, then vertex $0$ of edge
    %$e$ corresponds to the triangle vertices $\Delta_1\of0$
    %and $\Delta_2\of2$, and vertex $1$ of edge $e$ corresponds to
    %$\Delta_1\of2$ and $\Delta_2\of0$.
% \end{example}

Let $\tri$ be a $d$-dimensional triangulation.
The \emph{size} of $\tri$, denoted $|\tri|$, is the number of
simplices (i.e., $d$-faces) in $\tri$.
%Note that the total number of faces
%of \emph{any} dimension is at most $2^{d+1}|\tri|$, and so is
%linear in $|\tri|$ for fixed dimension $d$.
%
The \emph{dual graph} of $\tri$, denoted $\dual(\tri)$,
is the multigraph whose nodes correspond to simplices
and whose arcs correspond to identified pairs of facets.
%$\dual(\tri)$ has precisely $|\tri|$ nodes,
%each of degree $\leq d+1$.
% Loops may occur in $\dual(\tri)$ if two facets of the same simplex are
% identified, and parallel arcs may occur if different
% facets of some simplex $\Delta_i$ are identified with (different)
% facets of the same
% simplex $\Delta_j$.
See Figure~\ref{fig-dual} for an illustration.

The treewidth \cite{robertson86-algorithmic} of a graph or multigraph $G$
essentially measures how far $G$ is from
being a tree: any tree will have treewidth $1$,
and a complete graph on $n$ nodes will have treewidth $n-1$.
%Graphs of small treewidth are often easier to work
%with, as Courcelle's theorem (described below) so strikingly shows.
%The full definition is as follows.
%
%
More precisely, given a simple graph or multigraph $G$ with node set $V$,
a \emph{tree decomposition} of
$G$ consists of a (finite) tree $T$ and \emph{bags} $B_\tau \subseteq V$ for
each node $\tau$ of $T$ that satisfy the following constraints:
(i)~each $v \in V$ belongs to some bag $B_\tau$;
(ii)~for each arc of $G$, its two endpoints $v,w$ belong to some
common bag $B_\tau$; and
(iii)~for each $v \in V$, the bags containing $v$ correspond to a
(connected) subtree of $T$.
The \emph{width} of this tree decomposition is $\max |B_\tau|-1$, and the
\emph{treewidth} of $G$
is the smallest width of any tree decomposition of $G$, which
we denote by $\tw(G)$.

%Although computing treewidth is NP-complete, it is also fixed-parameter
%tractable: Bodlaender shows that for simple graphs of
%universally bounded tree\-width $\leq k$, one can compute tree
%decompositions of width $k$ in linear time \cite{bodlaender96-linear}.

%%%%%%%%%%%%%%%%%%%%%%%%%%%%%%%%%%%%%%%%%%%%%%%%%%%%%%%%%%%%%%%%%%%%%%%%

\section{Courcelle's theorem}

Monadic second-order logic, or MSO logic, is our framework for making
statements about triangulations.
%What we describe here is sometimes called \emph{extended} MSO logic,
%or $\mathit{MS}_2$ logic;
%this highlights the fact that we can access arcs directly through variables
%and sets, and not just indirectly through a binary relation on nodes.
%
Traditionally MSO logic is expressed in the framework of graph theory;
see a standard text such as \cite{flum06} for details.
Here we extend MSO logic to the setting of
$d$-dimensional triangulations, for fixed dimension $d \in \N$.
In this setting, we define MSO logic to support:
\begin{itemize}
    \item all of the standard boolean operations of
    propositional logic: $\wedge$ (and), $\vee$ (or), $\neg$ (negation),
    $\rightarrow$ (implication), and so on;

    \item for each $i=0,\ldots,d$, variables to represent
    $i$-faces of a triangulation, or sets of $i$-faces of a triangulation;

    \item the standard quantifiers from first-order logic:
    $\forall$ (the universal quantifier), and
    $\exists$ (the existential quantifier);
    % which may be applied to any of these variable types;

    \item the binary equality relation $=$,
    % which can be applied to any of these variable types,
    and the binary inclusion relation $\in$
    which relates $i$-faces to sets of $i$-faces;

    \item for each $i=0,\ldots,d-1$ and for each ordered sequence
    $\pi_0,\ldots,\pi_i$ of distinct integers from $\{0,\ldots,d\}$,
    a subface relation $\leq_{\pi_0 \ldots \pi_i}$.
\end{itemize}

% Note that sets are simply a convenient representation of unary relations
% on nodes and arcs.  A distinguishing feature of MSO logic is that
% we can quantify over unary relations (i.e., we can quantify
% over set variables as outlined above).

The relation $(f \leq_{\pi_0 \ldots \pi_i} s)$ indicates that
$f$ is an $i$-face of the triangulation,
$s$ is a simplex of the triangulation,
and that $f$ is identified with the subface of $s$ formed by the simplex
vertices $\pi_0,\ldots,\pi_i$, in a way that vertices $0,\ldots,i$ of
the face $f$ correspond to vertices $\pi_0,\ldots,\pi_i$ of the simplex $s$.

For example, recall the Klein bottle illustrated in Figure~\ref{fig-kb}.
Here the three edges $e,f,g$ satisfy the subface relations
$e \leq_{02} \Delta_1$,
$e \leq_{20} \Delta_2$,
$f \leq_{01} \Delta_1$,
$f \leq_{12} \Delta_1$,
$g \leq_{01} \Delta_2$,
$g \leq_{12} \Delta_2$.
%The triangulation has only one vertex (since all vertices of
%$\Delta_1$ and $\Delta_2$ are identified together); call this
%$v$.  Then $v$ satisfies all subface relations
%$v \leq_{0} \Delta_1$,
%$v \leq_{0} \Delta_2$,
%$v \leq_{1} \Delta_1$,
%$v \leq_{1} \Delta_2$,
%$v \leq_{2} \Delta_1$,
%$v \leq_{2} \Delta_2$.

We use the notation $\phi(x_1,\ldots,x_t)$ to denote an MSO formula
with $t$ free variables (i.e., variables not bound by
$\forall$ or $\exists$ quantifiers).
An \emph{MSO sentence} has no free variables at all.
If $\tri$ is a $d$-dimensional triangulation and $\phi$ is an
MSO sentence as above, then $\tri \models \phi$ indicates
that the interpretation of $\phi$ in the triangulation $\tri$
is a true statement.

%We prove three variants of Courcelle's theorem on triangulations,
%which relate to
%algorithms for (i)~\emph{decision} problems that test boolean conditions;
%(ii)~\emph{extremum} or optimisation problems; and
%(iii)~\emph{evaluation} or summation problems.

An \emph{MSO decision problem} is just an MSO sentence $\phi$.
Given a $d$-dimensional triangulation $\tri$ as input,
it asks whether $\tri \models \phi$.

A \emph{restricted MSO extremum problem} consists of
an MSO formula $\phi(A_1,\ldots,A_t)$
with free set variables $A_1,\ldots,A_t$,
and a rational linear function $g(x_1,\ldots,x_t)$.  Its interpretation
is as follows: given a $d$-dimensional triangulation $\tri$ as input,
we are asked to minimise $g(|A_1|,\ldots,|A_t|)$ over all sets
$A_1,\ldots,A_t$ for which $\tri \models \phi(A_1,\ldots,A_t)$,
where $|A_i|$ denotes the number of objects in the set $A_i$.

An \emph{MSO evaluation problem}
consists of an MSO formula $\phi(A_1,\ldots,A_t)$
with $t$ free set variables $A_1,\ldots,A_t$.
The input to the problem is a $d$-dimensional triangulation $\tri$,
together with $t$ weight functions
$w_1,\ldots,w_t \co F_0 \sqcup \ldots \sqcup F_d \to R$,
where $F_i$ denotes the set of all $i$-faces of $\tri$,
and $R$ is some ring or field.
The problem then asks us to compute one of the quantities
\[ \sum_{\tri \models \phi(A_1,\ldots,A_t)}
    \left\{
    \sum_{i=1}^t \sum_{x_i \in A_i} w_i(x_i)
    \right\}
    \quad\mbox{or}\quad
   \sum_{\tri \models \phi(A_1,\ldots,A_t)}
    \left\{
    \prod_{i=1}^t \prod_{x_i \in A_i} w_i(x_i)
    \right\}; \]
we refer to these two variants as \emph{additive} and
\emph{multiplicative} evaluation problems respectively.
For both problems, the outermost sum is over all solutions
$A_1,\ldots,A_t$ that satisfy $\phi$ on the triangulation $\tri$.

MSO evaluation problems should be thought of as generalised
counting problems: essentially,
we assign a value to each solution to some MSO formula,
and then sum these values over all solutions.
Counting problems themselves are simply multiplicative problems with
all weights $w_i=1$.

%Recall that the \emph{uniform cost measure}
%assumes that elementary arithmetic operations run in constant time;
%see a standard text such as \cite{aho75} for details.
%For MSO evaluation problems, we interpret the uniform cost measure
%to allow constant-time arithmetic operations over the ring or field $R$.

Our main result is the following:

\begin{theorem} \label{t-tri}
    For fixed dimension $d \in \N$,
    let $K$ be any class of $d$-dimensional triangulations whose dual
    graphs have universally bounded treewidth.  Then:
    \begin{itemize}
        \item
        For any fixed MSO sentence $\phi$,
        it is possible to test whether $\tri \models \phi$
        for triangulations $\tri \in K$ in time $O(|\tri|)$.
        \item
        For any restricted MSO extremum problem $P$,
        it is possible to solve $P$ for triangulations
        $\tri \in K$ in time $O(|\tri|)$ under the uniform cost measure.
        \item
        For any MSO evaluation problem $P$,
        it is possible to solve $P$ for triangulations
        $\tri \in K$ in time $O(|\tri|)$ under the uniform cost measure.
    \end{itemize}
\end{theorem}

In other words, solving any such problem is linear-time
fixed-parameter tractable in the treewidth of the dual graph.
By a result of Bodlaender \cite{bodlaender96-linear},
we do not need to supply an explicit tree decomposition of
$\dual(\tri)$ in advance.

In essence, the proof uses a series of constructions that
encode the full structure of a triangulation as a simple graph,
in a way that controls the growth of both the treewidth and the
input size.  From here we can invoke classical variants of
Courcelle's theorem from graph theory
\cite{arnborg91-easy,courcelle01-enumeration,courcelle87-context-free,courcelle90-rewriting}.

%The proof makes critical use of our definition of a $d$-dimensional
%triangulation, in that
%we only explicitly identify \emph{facets} of $d$-simplices, and all
%lower-dimensional face identifications just follow from this.
%The proof does not work for more general simplicial complexes (where
%lower-dimensional faces can be independently identified), but we
%note that our setting here covers all reasonable definitions of a
%triangulated \emph{manifold}.
%Moreover, it covers
%structures beyond simplicial complexes,
%such as the highly efficient
%\emph{one-vertex triangulations} % \cite{jaco03-0-efficiency}
%and \emph{ideal triangulations} % \cite{thurston78-lectures}
%favoured by many 3-manifold topologists.

%%%%%%%%%%%%%%%%%%%%%%%%%%%%%%%%%%%%%%%%%%%%%%%%%%%%%%%%%%%%%%%%%%%%%%%%
%
%   Applications
%
%%%%%%%%%%%%%%%%%%%%%%%%%%%%%%%%%%%%%%%%%%%%%%%%%%%%%%%%%%%%%%%%%%%%%%%%

\section{Applications} \label{s-app}

Our first application is in discrete Morse theory, which
offers a combinatorial way to study the
``topological complexity'' of a triangulation.  The idea is to
effectively quarantine the topological content of a triangulation
into a small number of ``critical faces''; the remainder of the
triangulation then becomes ``padding'' that is topologically unimportant.
A key problem in this area is to find an
\emph{optimal Morse matching}, where the number of critical faces is
as small as possible.  Solving this problem yields important topological
information, and has a number of practical applications.

In dimension $d=3$ the problem of finding an optimal Morse function
for a given $d$-dimensional triangulation is NP-complete
\cite{joswig06-morse}, but linear-time fixed-parameter tractable
in the treewidth of the dual graph \cite{burton13-morse}.
Here we generalise the latter result to arbitrary dimensions:

\begin{theorem} \label{t-morse}
    For fixed dimension $d \in \N$ and
    any class $K$ of $d$-dimensional triangulations whose dual
    graphs have universally bounded treewidth,
    we can find an optimal Morse matching
    for triangulations $\tri \in K$ in time $O(\tri)$
    under the uniform cost measure.
\end{theorem}

%%%%%%%%%%%%%%%%%%%%%%%%%%%%%%%%%%%%%%%%%%%%%%%%%%%%%%%%%%%%%%%%%%%%%%%%

Our second application is for
the Turaev-Viro invariants, an infinite family of topological
invariants of 3-manifolds \cite{turaev92-invariants}.
For every triangulation $\tri$ of a closed 3-manifold, there is an invariant
$|\tri|_{r,q_0}$ for each integer $r \geq 3$
and each $q_0 \in \C$ for which $q_0$ is a $(2r)$th root of unity
and $q_0^2$ is a primitive $r$th root of unity.
The value of $|\tri|_{r,q_0}$ depends only upon the topology of
the underlying 3-manifold.

% Naive running time:
% v <= n+2 (B.B., JCTA 2011)
% e = n+v <= 2n+2
% time: (r-1)^(2n+2)

The Turaev-Viro invariants can be expressed as sums over combinatorial
objects on $\tri$, and so (unlike many other 3-manifold invariants)
lend themselves well to computation.
Moreover, they have proven extremely powerful in practical software settings
for distinguishing between different 3-manifolds.
However, they have a major drawback: computing $|\tri|_{r,q_0}$
requires time $O(r^{2|\tri|} \times \mathrm{poly}(|\tri|))$
under existing algorithms, and so is feasible
only for small $|\tri|$ and/or $r$.
Here we show that we can do much better for small treewidth triangulations:

\begin{theorem} \label{t-tv}
    For any fixed integer $r \geq 3$
    and any class $K$ of closed $3$-manifold triangulations
    whose dual graphs have universally bounded treewidth,
    we can compute any Turaev-Viro invariant
    $|\tri|_{r,q_0}$ for any closed 3-manifold triangulation
    $\tri \in K$ in time $O(\tri)$
    under the uniform cost measure.
\end{theorem}

Although ``treewidth of the dual graph'' seems an artificial parameter,
it is natural and useful for
3-manifold triangulations---here many common constructions
are conducive to small treewidth even when the input size is large.
For example:
Dehn fillings do not increase treewidth when performed
``efficiently'' by attaching layered solid tori;
``canonical'' triangulations of arbitrary Seifert fibred spaces
over the sphere have treewidth bounded by just two;
and building a complex 3-manifold triangulation from smaller
blocks with ``narrow'' $O(1)$-sized connections (e.g., via
JSJ decompositions) can also keep treewidth small.

%%%%%%%%%%%%%%%%%%%%%%%%%%%%%%%%%%%%%%%%%%%%%%%%%%%%%%%%%%%%%%%%%%%%%%%%
%
%   Bibliography
%
%%%%%%%%%%%%%%%%%%%%%%%%%%%%%%%%%%%%%%%%%%%%%%%%%%%%%%%%%%%%%%%%%%%%%%%%

\bibliographystyle{amsplain}
{\small
\bibliography{pure}
}

\end{document}

\maketitle

\begin{abstract}
A triangulation of a closed $d$-manifold is said to be {$r$-stacked} if it is the boundary of a triangulation of a (homology) $(d+1)$-manifold having no interior faces of dimension $\leq d-r$.
%The notion of $r$-stackedness was introduced by McMullen and Walkup in 1971 for triangulations of spheres
%as a generalization of stacked polytopes.
In this abstract, 
we introduce several interesting properties of $r$-stacked triangulated manifolds.
\end{abstract}

\section*{Introduction}
Recently, the $r$-stacked property of triangulated spheres has been of interest in the study of face numbers of simplicial complexes.
A triangulated $d$-sphere is said to be {\em $r$-stacked} if it is the boundary of a triangulated (homology) $(d+1)$-ball
having no interior faces of dimension $\leq d-r$.
This $r$-stacked property can be naturally extended to triangulations of manifolds,
and it was found that $r$-stacked triangulated $d$-manifolds have many interesting properties when $r < \frac d 2$.
In this abstract, we summarize known nice properties of $r$-stacked triangulated manifolds.

We first introduce necessary notations.
Let $\Delta$ be a (finite abstract) simplicial complex.
We say that a simplicial complex $\Delta$ is a \emph{triangulation} of a topological space $X$ if its geometric realization is homeomorphic to $X$.
We are interested in triangulations of manifolds, but we actually consider slightly larger class of simplicial complexes.
Fix a field $\kk$.
For a simplicial complex $\Delta$ and its face $F \in \Delta$,
the {\em link of $F$ in $\Delta$} is the simplicial complex
$$\lk_\Delta(F)=\{ G \in \Delta: F \cup G \in \Delta \mbox{ and } F \cap G = \emptyset\}.$$
A simplicial complex $\Delta$ of dimension $d$ is said to be a {\em homology $d$-sphere}
if, for all faces $F \in \Delta$ (including the empty face $\emptyset$),
one has $\beta_i(\lk_\Delta(F))=0$ for $i \ne d-\#F$ and $\beta_{d-\#F}(\lk_\Delta(F))=1$,
where $\beta_i(\Delta)=\dim_\kk \widetilde H_i(\Delta;\kk)$ is the $i$th {\em Betti number} of $\Delta$ (w.r.t.\ $\kk$).
A simplicial complex is said to be {\em pure} if all its facets have the same dimension.
A {\em homology $d$-manifold (without boundary)} is a pure $d$-dimensional simplicial complex
all of whose vertex links are homology spheres.
A pure $d$-dimensional simplicial complex $\Delta$ is said to be a {\em homology $d$-manifold with boundary} if it satisfies
\begin{itemize}
\item[(i)] for all $\emptyset \ne F \in \Delta$,
$\beta_i(\lk_\Delta(F))$ vanishes for $i \ne d - \#F$ and is equal to $0$ or $1$ for $i = d - \#F$.
\item[(ii)] the {\em boundary} $\partial \Delta=\{F \in \Delta: \beta_{d-\#F}(\lk_\Delta(F))=0\} \cup \{ \emptyset \}$ of $\Delta$
is a homology $(d-1)$-manifold without boundary.
\end{itemize}
A {\em homology $d$-ball} is an acyclic homology $d$-manifold such that $\partial \Delta$ is a homology $(d-1)$-sphere.
A triangulation of a topological manifold is an example of a homology manifold.
Also, a triangulation of a $d$-sphere (resp.\ $d$-ball) is a homology $d$-sphere (resp.\ $d$-ball).

\section{Examples of stacked polytopes and manifolds}

We say that a homology $d$-manifold $\Delta$ with boundary is {\em $r$-stacked} if it has no interior faces (namely, faces which are not in $\partial \Delta$) of dimension $< d-r$.
Also, a homology manifold without boundary is said to be {\em $r$-stacked} if it is the boundary of an $r$-stacked homology manifold with boundary.
Although we defined $r$-stacked manifolds with boundary, we are mainly interested in homology manifolds without boundary.
So we often omit `without boundary' when it is clear.
We also assume that all homology manifolds are connected.
Here we give two examples of $1$-stacked triangulated manifolds.

\begin{example}
A $d$-polytope is said to be {\em stacked} if it can be obtained from a simplex by adding a pyramid over a facet repeatedly. It is not hard to see that the boundary complexes of stacked polytopes are exactly 
$1$-stacked spheres.
\medskip

\begin{center}
\includegraphics[width=100mm]{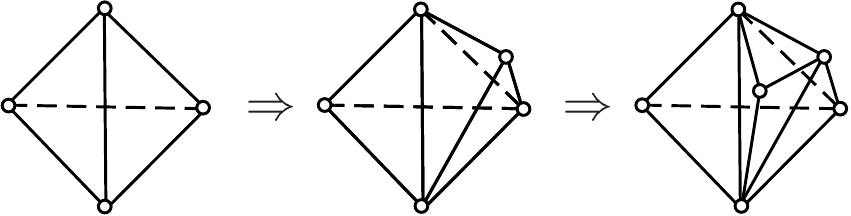}\\
construction of stacked polytopes
\end{center}

An important property of stacked polytopes is that it gives a lower bound of face numbers of simplicial polytopes for a fixed dimension and a fixed number of vertices.
See \cite{Ba1,Ba2,Ka}.
\end{example}

\begin{example}[K\"uhnel--Lassmann construction \cite{Ku,KL}]
\label{ex1}
Let $d, n$ be integers with $ n \geq 2d-1$ and let
$K_{d,n}$ be the simplicial complex on $[n]=\{1,2,\dots,n\}$ generated by the facets
$$\{ \{i,i+1,\dots,i+d-1\}: i=1,2,\dots,n\},$$
where $i+k$ means $i+k-n$ if $i+k >n$.
Then $K_{d,n}$ is a combinatorial manifold such that $\partial K_{d,n}$
triangulates either $S^{d-3} \times S^1$ or a non-orientable $S^{d-3}$-bundle over $S^1$ (generalized Klein bottle).
The interior faces of $K_{d,n}$ are faces that contain a face of the form $\{i,i+1,\dots,i+d-2\}$ (mod $n$),
so $K_{d,n}$ and $\partial K_{d,n}$ are $1$-stacked triangulations.

An important property of these triangulations is that they give vertex minimal triangulations of a sphere-bundle over a circle when $n=2d-1$.
For example, $\partial K_{4,7}$ is Cs\'asz\'ar's vertex minimal $7$-vertex triangulation of $S^1 \times S^1$.
We will explain later why the stackedness property appears in minimal triangulations.
\end{example}

\section{Properties of stacked triangulations}

In this section, we list known nice properties of $r$-stacked triangulated manifolds.

\subsection{Construction of $r$-stacked triangulated manifolds with boundary}
One of the most interesting properties of stacked triangulations is that
one can construct the unique $r$-stacked triangulated $d$-manifold with boundary from an $r$-stacked triangulated $(d-1)$-manifold without boundary when $r$ is small enough. 
For a simplicial complex $\Delta$ on $V$,
define the simplicial complex
$$\Delta(r)=\{ F \subset V: \mbox{ any subset $G \subset F$ with $\# G \leq r+1$ is contained in }\Delta\}.$$ 
The following result was proved by Bagchi and Datta for simplicial polytopes and proved in \cite{MN1} for homology spheres. See \cite[Theorem 2.3]{MN1}.

\begin{theorem}
Let $1 \leq r \leq \frac {d} 2$ and let $\Delta$ be an $r$-stacked homology $d$-sphere. 
If $\Sigma$ is an $r$-stacked homology $(d+1)$-ball with $\partial \Sigma=\Delta$ then $\Sigma=\Delta(r)$.
\end{theorem}

For triangulated manifolds, a slightly weaker statement was proved in \cite[Theorem 2.20]{BD} and in \cite[Theorem 4.2]{MN2} independently.

\begin{theorem}
\label{4.3}
Let $1 \leq r \leq \frac {d-1} 2$ and
$\Delta$ an $r$-stacked homology $d$-manifold without boundary.
If $\Sigma$ is an $r$-stacked homology $(d+1)$-manifold with $\partial \Sigma=\Delta$ then $\Sigma=\Delta(r+1)$.
\end{theorem}

Note that these theorems also say that the $r$-stacked homology $d$-manifold $\Sigma$ with $\partial \Sigma=\Delta$ is unique under these assumptions.
A more interesting aspect of the above theorems is that if one happens to know that $\Delta$ is an $r$-stacked triangulated $d$-manifold, then the operation $\Delta \to \Delta(r+1)$ gives rise to a homology $(d+1)$-manifold.

\subsection{Local criterion}
We say that a homology $d$-manifold without boundary is {\em locally $r$-stacked}
if all its vertex links are $r$-stacked.
This notion was studies by Walkup \cite{Wa} and Kalai \cite{Ka} when $r=1$
and the class of locally $1$-stacked triangulated manifolds are known as Walkup's class.
The locally $r$-stacked property with $r >1$ also appeared in \cite{Ef}.
It is easy to see that $r$-stacked manifolds are locally $r$-stacked.
The next statement proved by Bagchi--Datta \cite[Theorem 2.19]{BD} and in \cite[Theorem 4.6]{MN2} shows that they are equivalent when $r$ is small. 

\begin{theorem}
\label{4.4}
Let $1 \leq r< \frac {d-1} 2$.
Then a homology $d$-manifold without boundary is $r$-stacked if and only if it is locally $r$-stacked.
\end{theorem}

Theorem \ref{4.4} fails for $r= \frac {d-1} 2$.
For example, it is known that the boundary complex of a cyclic $4$-polytope is locally $1$-stacked but not $1$-stacked.
This implies that, for $3$-manifolds, we cannot study the $1$-stacked property by just considering vertex links.

\subsection{Stacked property and face numbers}
The $r$-stacked property was considered in the study of face numbers
of simplicial polytopes.
Let $\Delta$ be a simplicial complex of dimension $d-1$.
Let $f_i(\Delta)$ be the number of $i$-dimensional faces of $\Delta$.
The vector $f(\Delta)=(f_{-1}(\Delta),\dots,f_{d-1}(\Delta))$ is called the {\em $f$-vector} of $\Delta$.
Also, the {\em $h$-vector} $h(\Delta)=(h_0(\Delta),h_1(\Delta),\dots,h_d(\Delta))$ of $\Delta$ is
defined by the relation
$$
\sum_{i=0}^d h_i(\Delta) t^i = \sum_{i=0}^d f_{i-1}(\Delta) t^i (1-t)^{d-i}.
$$
We say that a simplicial $d$-polytope is $r$-stacked if it can be (geometrically) triangulated without introducing faces of dimension $d<r$.
Note that this definition coincides with the $r$-stackedness of its boundary complex when $r \leq \frac {d-1} 2$.
For simplicial polytopes, the following characterization of the $r$-stacked property is known.

\begin{theorem}[Generalized lower bound theorem]
\label{glbc}
Let $P$ be a simplicial $d$-polytope.
\begin{itemize}
\item[(i)] (Stanley \cite{St}) $h_r(P) \leq h_{r+1}$ for all $r \leq \frac {d-1} 2$.
\item[(ii)] (McMullen--Walkup \cite{MW}, Murai--Nevo \cite{MN1}) for $r < \frac {d-1} 2$,
one has $h_r(P)= h_{r+1}(P)$ if and only if $P$ is $r$-stacked.
\end{itemize}
\end{theorem}

The special case of the above theorem when $r=1$ is known to be equivalent to Barnette's Lower bound theorem \cite{Ba1,Ba2}.
It is a natural question to ask if Theorem \ref{glbc} can be extended to triangulations of spheres.
Unfortunately, this seems to be a difficult problem since the proof of the theorem depends on the hard Lefschetz theorem for projective toric varieties.
On the other hand, the following conjecture is suggested.

\begin{conjecture}[{Generalized lower bound conjecture for triangulated manifolds}]
\label{5.1}
Let $\Delta$ be a connected triangulated $(d-1)$-manifold without boundary. Then
\begin{itemize}
\item[(i)] (Kalai) $h_{r+1}(\Delta) \geq h_r(\Delta)+ \binom {d+1} {r+1} \sum_{j=1}^{r+1} (-1)^{r+1-j} \beta_{j-1}(\Delta)$ for $r=1,2,\dots,\lfloor \frac d 2 \rfloor$.
\item[(ii)] (Bagchi--Datta) if equality holds for some $r< \frac d 2 -1$ in (i)
then $\Delta$ is locally $r$-stacked.
\end{itemize}
\end{conjecture}

The case $r=1$ was solved by Novik--Swartz \cite{NS1}.
The conjecture is also known to hold when $\Delta$ is orientable and all the vertex links of $\Delta$ are simplicial polytopes \cite{NS2,MN2}.

\subsection{Vertex minimal triangulations}
Stacked triangulations sometimes appear as minimal triangulations.
Here we explain a reason.
Let $\Delta$ be a homology $d$-manifold without boundary.
It was conjectured by K\"uhnel and proved by Novik--Swartz that
the inequality
\begin{eqnarray}
\label{1-1}
{{d+2} \choose 2 }\beta_1(\Delta) 
\leq {f_0(\Delta)-d-1 \choose 2}
\end{eqnarray}
holds.
A triangulated $d$-manifold satisfying \eqref{1-1} with equality is called a {\em tight-neighborly} triangulation.
Clearly a tight-neighborly triangulation must be a vertex minimal triangulation.
The following result follows from the works of Novik--Swartz \cite[Theorem 5.3]{NS1}, Bagchi--Datta \cite[Theorem 2.24]{BD} and Burton--Datta--Singh--Spreer \cite[Theorem 1.2]{BDSS}.

\begin{theorem}
Tight neighborly triangulated manifolds are $1$-stacked.
\end{theorem}

An $r$-stacked analogue of the above theorem is expected to be true.
It is known that all the vertex minimal triangulations of $S^2 \times S^3$ listed in \cite{Lu} are $2$-stacked. It would be interesting to find a result that explains this fact.

\subsection{Topological restriction}
It was proved by Kalai \cite{Ka} that if $\Delta$ is a locally $1$-stacked triangulated manifold then it
is obtained from the boundary of a simplex by repeating the following three operations: (i) stellar subdivision of a facet (i) combinatorial handle addition (iii) taking connected sums of objects obtained from the first two operations.
This gives a strong restriction to topological types of $1$-stacked triangulated $d$-manifolds with $d \geq 4$.
A natural question is ``what can be said about the topological type of $r$-stacked triangulated $d$-manifolds with $r < \frac {d-1} 2$?''
At the moment, we only have information on homology groups. See \cite[Remark 2.22]{BD} and \cite[Theorem 4.4]{MN2}.
Also, we are not sure that, if $\Delta$ is an $r$-stacked homology $d$-manifold with $r \leq \frac {d-1} 2$, then the topological type of $\Delta(r+1)$ in Theorem~2.2 can be determined from the topological type of $\Delta$.

%\begin{theorem}
%Let $ 1 \leq r < \frac d 2$ and let $\Delta$ be an $(r-1)$-stacked homology $(d-1)$-manifold.
%Then
%\begin{itemize}
%\item[(i)] (\cite[Theorem 4.4]{MN2}) $\beta_i(\Delta)=0$ for $r \leq i \leq d-1 -r$. 
%\item[(ii)] (Bagchi-Datta) $H_{r-1}(\Delta)$ is torsion free.
%\end{itemize}
%\end{theorem}

%\noindent
%\textbf{Acknowledgments}:

\maketitle

\begin{abstract}
Normal surface theory is the study of surfaces embedded in a triangulation of a 3-manifold.  The surfaces inherit a combinatorial structure from the triangulation which can be exploited to prove finiteness and algorithmic results about the manifold itself.   This is the basis of Haken's algorithm to recognize the unknot, Rubinstein and Thompson's algorithm to recognize the 3-sphere, as well as Matousek, Sedgwick, Tancer and Wagner's recent algorithm to determine whether a 3-manifold, hence a 2-complex, embeds in the 3-sphere. We give an overview of normal surface theory, its application to recognition and embedding, and the refinements that make the embedding result possible.
\end{abstract}

\section*{Algorithms for 3--Manifolds}

For each of the following decision problems an affirmative outcome is witnessed by a surface embedded in the 3--manifold:

\begin{theorem*}[Unknot Recognition -  \cite{Haken:Theorie-der-Normalflachen:-Ein-Isotopiekriterium-fur-den-Kreisknoten-1961}]   There is an algorithm to recognize the unknot:   Let $K$ be a polygonal loop in $S^3$.  Then there is an algorithm to determine whether $K$ is an unknot, i.e., whether $K$ bounds a disk in $S^3$.
\end{theorem*}

It is straightforward to triangulate the {\it exterior of $K$}, $X = S^3 - N(K)$.  If the 3--manifold $X$ contains an embedded disk meeting the boundary of $X$ in an essential curve, this disk serves as a witness that $K$ is an unknot.

\begin{theorem*}[$S^3$ Recognition -  \cite{Rubinstein:AlgorithmRecongnizing3Sphere-1995},  \cite{Thompson:ThinPositionrecognitionProblemS3-1994}]  Let $X$ be a triangulated 3--manifold.  There is an algorithm to determine whether $X$ is (homeomorphic to) the 3--sphere, $S^3$.
\end{theorem*}

The 3--sphere is the union of two balls glued along their boundary.   Any embedded 2--sphere known to separate $X$ into a pair of balls is a witness to the fact that $X$ is $S^3$.

\begin{theorem*}[Embeddability in $S^3$ -  \cite{matousekSedgwickTancerWagnerEmbed3}] Let $X$ be a triangulated 3--manifold.  There is an algorithm to determine whether $X$ embeds in $S^3$.
\end{theorem*}

Generically, if $X$ embeds in $S^3$ then $X$ embeds in $S^3$ so that its complement is a collection of thickened graphs called {\it handlebodies} \cite{Fox:On-the-imbedding-of-polyhedra-in-3-space-1948}.  Then there is an embedded sphere in $S^3$ that meets the handlebodies only in (essential) disks.  Such a sphere meets $X$ in a planar surface that is said to be {\it meridional}.     By {\it filling}, attaching thickened disks to $X$ along the surfae's meridional boundary curves, we obtain a manifold $X'$ that also embeds in $S^3$ but whose complementary handlebodies correspond to a graph with fewer edges (at least one was cut).   Thus, a meridional planar surface witnesses a step of an inductive proof that $X$ embeds in $S^3$ (other reductions are also required).

\section*{Normal Surface Theory}

In this section we describe how normal surface theory is applied to these problems.  We leave aside many details and focus on the mechanics of the method.   The reader is referred to \cite{Jaco:Algorithms-for-the-complete-decomposition-of-a-closed-3-manifold-1995} for a more complete treatment.

A {\it normal surface} is an embedded surface that meets each tetrahedron of the triangulation in a collection of disks,  each disk is either a triangle that cuts off one of the four corners of the tetrahedron, or, a quadrilateral separating a pair of edges of the tetrahedron. Each  normal surface $N$ is uniquely determined by a vector that counts the number of pieces of each of the possible types, $\vv(N) \in \mathbb N^{7t}$, where $t$ is the number of tetrahedra.

The vector $\vv(N)$ also satisfies a set of at most $6t$ {\it matching equations}  that ensure that the pieces in neighboring tetrahedra match up on the face they share.  Each matching equation has the form $v_i+v_j=v_k+v_l$, setting equal the sum of the counts of a triangle and quadrilateral from one side to the sum of the counts of a triangle and quadrilateral on the other.

For each normal surface $N$, the vector $\vv(N)$ is thus a solution to a set of linear equations with integer coefficients.  A normal surface is {\it fundamental} if its vector cannot be written as a non-trivial positive sum of other solutions.  Each fundamental belongs to the constructable {\it minimal Hilbert basis} for the system.  Their number and the weight of each are bounded by  functions of $t$ \cite{hass-lagarias-pippenger}.

In an abuse of notation, we write each normal surface (rather than its vector) as a sum of fundamental surfaces $$N = \sum k_i F_i,~~ k_i \geq 0.$$

This sum has several desirable properties, notably:

\begin{enumerate}
\item Euler characteristic is additive: $\chi(N) = \sum k_i \chi(F_i)$
\item Weight, the number of intersections with edges, is additive: $w(N) = \sum k_i ~w(F_i)$

\item Length, the number of intersections of the surface's boundary with edges, is additive: $\ell(N) = \sum k_i ~\ell(F_i)$

\end{enumerate}

Another essential ingredient, mostly omitted here, is that under achievable side conditions,  topological properties of $N$ (incompressibility, essentiality) are also held by the summands \cite{Jaco:Algorithms-for-the-complete-decomposition-of-a-closed-3-manifold-1995}.   This requires the {\it Haken sum},  an interpretation of the vector sum as a geometric sum of surfaces.

\subsection*{Unknot Recognition} If $X = S^3-N(K)$ contains an essential unknotting disk $D$, then that disk is isotopic to a normal surface.  Moreover, using the geometric interpretation and additivity of Euler characteristic, it can be shown that one of the fundamentals is an essential unknotting disk.

To decide whether $K$ is an unknot, triangulate the complement of $K$, $X = S^3 - N(K)$, and construct the fundamental $\{F_i\}$ solutions.     The loop $K$ is an unknot if and only if some $F_i$ is an unknotting disk, a straightforward calculation.

\subsection*{$S^3$ Recognition}

Spheres in $S^3$ are not essential and one cannot expect to find a useful normal witness.   But, using Gabai's notion of {\it thin position} \cite{gabai1983}, a sphere can be isotoped to be {\it almost normal} in the triangulation \cite{Thompson:ThinPositionrecognitionProblemS3-1994}, that is, normal everywhere except for one exceptional piece that is either an octagon or a tube.  Almost normal surfaces are also represented by vectors, albeit with more coordinates, and the mechanics of matching equations and computing fundamental solutions remain the same.  Again, additivity of Euler characteristic is crucial in concluding that  spheres ($\chi=2$) of interest occur among the fundamentals.

Deciding whether $X$ is $S^3$ rests on decomposing $S^3$ along normal spheres and deciding whether {\it almost normal} spheres occur among the fundamental solutions to the matching equations for certain types of pieces.

\subsection*{Embeddability in $S^3$}

Generically, when $X$ embeds in $S^3$, its triangulation contains an almost normal meridional planar surface $P$, using \cite{Fox:On-the-imbedding-of-polyhedra-in-3-space-1948}, \cite{Li:Thin-position-and-planar-surfaces-for-graphs-in-the-3-sphere-2010}, and \cite{bachmanDerbyTalbotSedgwick}.  But while it can be written as the sum of fundamentals $P = \sum k_i \chi(F_i)$, its Euler characteristic is negative and without bound.   We therefore have no bound on the coefficients $k_i$ and no hope of finding such a surface among the fundamentals.

Fortunately, an alternate strategy, the {\it average length estimate} \cite{Jaco:Finding-planar-surfaces-in-knot--and-link-manifolds-2009}, is available.    {\it If all summands have negative Euler characteristic}, then the total length of $P$ is bounded by  $\ell(P) \leq -\chi(P)~ \ell_{\max}$, where $\ell_{\max}$ is the maximum length realized among the fundamentals.    If $P$ has $b$ boundary components, then $\chi(P)=2-b$ and the average length of a boundary component of $P$ is indeed bounded

$$\ell(\textit{meridan})\leq\frac {\ell(P)} b   \leq \frac{ \ell(P)}{-\chi(P)}\leq \ell_{\max}$$

However, while fundamental summands of positive Euler characteristic can be ruled out by passing to a 0-efficient triangulation \cite{Jaco:0-efficient-triangulations-of-3-manifolds-2003}, and tori and Klein bottles have zero length and can be ignored, summands that are annuli and M\"obius summands are expected and problematic.   The desired result is recovered by proving that the coefficients of annuli and M\"obius bands can be taken to be sufficiently bounded in terms of $\chi(P)$.  An increased bound on meridian length is obtained that depends only on the number of tetrahedra.

The generic case of the algorithm is then to fill along all ``short'' curves and inductively determine whether the resulting manifold embeds in $S^3$.  The manifold embeds if and only one of the filled manifolds does.  Other reductions are also required.

%
%A crucial step in the argument is establishing another desirable property of normal summation:
%
%$$g(P,\alpha) = \sum k_i ~g(F_i,\alpha),$$
%
%Here $g(\cdot,\alpha)$ is the geometric intersection number, the minimal number of intersections up to isotopy, and, $\alpha$ is a particular normal curve in the boundary of the manifold $X$.    We describe below an extension of normal surface theory that make this possible.
%
% The algorithm will fill along all short boundary curves, and induct.
%
%prove that some meridional boundary curve of $P$ is ``short'' and thus itself a suitable witness.

\section*{Normal Curves}

The boundary of an (almost) normal surface is a normal curve in the triangulated boundary of the manifold.   A {\it normal curve} is an embedded curve that meets each face of a triangulation in a collection of {\it normal arcs}, each arc joining distinct edges of the face.  The remaining discussion is mainly restricted to normal curves, although we expect the definitions and concepts to be useful in the context of normal surfaces.

The {\it geometric intersection number}, $i(\alpha,\beta)$, between a pair of curves $\alpha$ and $\beta$ is the minimal number of intersections up to isotopy of the curves.  Several refinements to normal curve/surface theory are introduced in \cite{matousekSedgwickTancerWagnerEmbed3} that improve the behaviour of normal curves with respect to geometric intersection number.

\subsection*{Tight and Snug Normal Curves}
 
If a surface has $f$ faces, each normal curve $\alpha$ is represented by a vector $\vv(\alpha) \in \mathbb N^{3f}$ which counts the number of each of the normal arc types.   The usual measure of complexity of a normal curve is its length $\ell(\alpha)$, the number of intersections of the curve with the edges of the triangulation.   We refine the notion by defining complexity to be a pair consisting of its length and its normal vector,
$$\compl(\alpha):=(\ell(\alpha),\vv(\alpha)).$$

Order complexities lexicographically, assuming an arbitrary but fixed ordering of the normal arcs types.   A curve is {\it essential} if it does not bound a disk in the surface.   A curve $\alpha$ is {\it tight} if it minimizes complexity over all curves to which it is isotopic. We obtain several properties for tight curves which do not hold in general for least length curves:

\begin{lemma}[\cite{matousekSedgwickTancerWagnerEmbed3}]
A tight essential curve is normal and unique up to normal isotopy.
\end{lemma}

A pair of normal curves is {\it snug} if their geometric intersection number is realized.   Tight curves are automatically snug:

\begin{lemma}[\cite{matousekSedgwickTancerWagnerEmbed3}] Let $\alpha$ and $\beta$ be tight essential normal curves.  Then $\alpha$ and $\beta$ are, after a normal isotopy, snug.
\end{lemma}

The summands of a tight essential curve are tight and essential:

\begin{lemma}[\cite{matousekSedgwickTancerWagnerEmbed3}]
Suppose that a tight essential normal curve is a normal sum $\alpha + \beta$.  Then $\alpha$ and $\beta$ are tight, essential, and after a normal isotopy, snug.
\end{lemma}

\subsection*{Marked Triangulations}

Finally we extend normal curve/surface theory to support additivity of geometric intersection numbers with a pre-determined normal curve.   A {\it marked triangulation} is a pair $(\T,M)$ consisting of a triangulation $\T$ along with a marking $M$, a finite set of points along the edges of $\T$. An arc is {\it $M$-normal} if it is normal in the triangulation  and disjoint from $M$.  An embedded curve is {\it $M$-normal} if it is the union of $M$-normal arcs.

While a marking increases both the number of coordinates and the number of matching equations the process of computing fundamentals is unchanged.  The expert should note that a marking introduces additional compatibility classes, even for normal curves.

A {\it fence} is an embedded curve that is normal and which meets the edges only at marked points.  Addition of $M$-normal curves behaves well with respect to fences.   In particular, geometric intersection number with a fence is additive:

\begin{theorem*}[\cite{matousekSedgwickTancerWagnerEmbed3}]
\label{propMSummands}
Let $\mu$ be a fence that is a tight essential curve (w.r.t.\ the unmarked
triangulation). Suppose that a sum  $\alpha+\beta$  of $M$-normal curves is tight, essential and snug with $\mu$.  Then
\begin{enumerate}

\item[\rm(1)] $\alpha$ and $\beta$ are both snug with respect to $\mu$;

 \item[\rm(2)] $i(\alpha+\beta,\mu) = i(\alpha,\mu) + i(\beta,\mu)$ where $i(.,.)$
is the geometric intersection number;

\end{enumerate}
\end{theorem*}

This theorem is applied for an embedding of $X$ in $S^3$ to obtain bounds on annulus and M\"obius band summands of the meridional planar surface.  In that case, the fence is the boundary of a maximal collection of essential annuli in $X$.

\bibliographystyle{alpha}
\bibliography{Postnikov,3Manifolds,SOCG}

\maketitle

\begin{abstract}
There are infinitely many triangulations of a given 3-manifold (here
we restrict to one-vertex triangulations for closed manifolds, and
ideal triangulations for manifolds with boundary). For each manifold $M$
we form a graph $\mathfrak{T}(M)$ whose vertices are the triangulations of $M$, and
for which two vertices are connected if the triangulations are related
by a Pachner 2-3 move. Matveev and Piergallini independently show that
for each manifold, $\mathfrak{T}(M)$ is connected (other than for triangulations
consisting of a single tetrahedron). However, very little else is
known about the structure of $\mathfrak{T}(M)$.

There are many useful properties a triangulation can have, for example
geometric triangulations, triangulations with angle structures, 0- and
1-efficient triangulations and triangulations with essential edges.
Almost nothing is known about the subgraphs of $\mathfrak{T}(M)$ corresponding to
these kinds of triangulation. I will survey these properties and the
relations between them, and say something about how we can start to
investigate their connectivity in $\mathfrak{T}(M)$.
\end{abstract}

\noindent
\textbf{Acknowledgement:} Many of the ideas in this talk have come from discussions with Craig D. Hodgson and J. Hyam Rubinstein.

\section{Pachner moves}

The set of triangulations $\mathfrak{T}(M)$ of a 3-manifold $M$ is still a poorly understood object. In comparison to similar objects, such as the set of Heegaard splittings of a 3-manifold, or the curve complex, almost nothing is known about the large scale properties of the set of triangulations.

We are mostly concerned with 1-vertex triangulations in the case of a closed 3-manifold, or ideal triangulations in the case of a 3-manifold with boundary, and it is this collection of triangulations that we denote by $\mathfrak{T}(M)$. We refer to the collection of triangulations allowing any number of vertices (in the case of a closed manifold) or finite vertices (in the case of a manifold with boundary) as $\overline{\mathfrak{T}}(M)$.

\begin{figure}[htb]
%\labellist
%\small\hair 2pt
%\pinlabel 1--4 at 145 93
%\pinlabel 2--3 at 491 93
%\endlabellist
\centering
\includegraphics[width=0.8\textwidth]{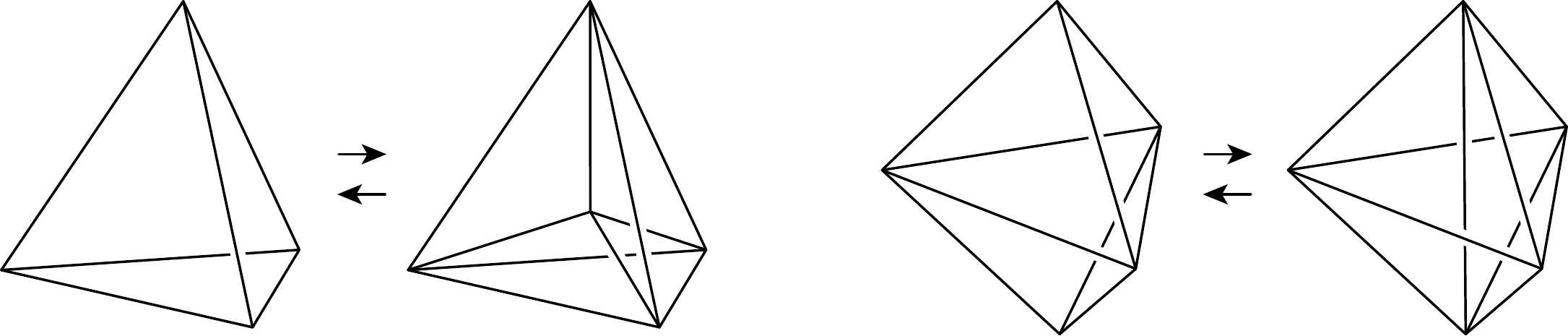}
\caption{The four Pachner moves in 3 dimensions.}
\label{Pachner moves}
\end{figure}

A natural structure on $\overline{\mathfrak{T}}(M)$ is that of the Pachner graph. A \emph{Pachner move} modifies a triangulation in a local fashion, by replacing a collection of simplices with another collection with the same boundary. In 3 dimensions, there are four such moves, the 1--4, 2--3, 3--2 and 4--1 moves. See Figure \ref{Pachner moves}. The Pachner graph then has a vertex for each triangulation of $\overline{\mathfrak{T}}(M)$ and an edge when two such triangulations are related by a Pachner move.

It is a result of Pachner \cite{pachner} that the Pachner graph is connected.
Matveev \cite{matveev_2-3_connectivity_paper, matveev} and Piergallini  \cite{piergallini} independently showed that $\mathfrak{T}(M)$ is connected using only the 2--3 and 3--2 moves (other than for those triangulations involving only a single tetrahedron). 
\section{Properties of triangulations}\label{properties}

\begin{figure}[htb]
\centering
\includegraphics[width=0.88\textwidth]{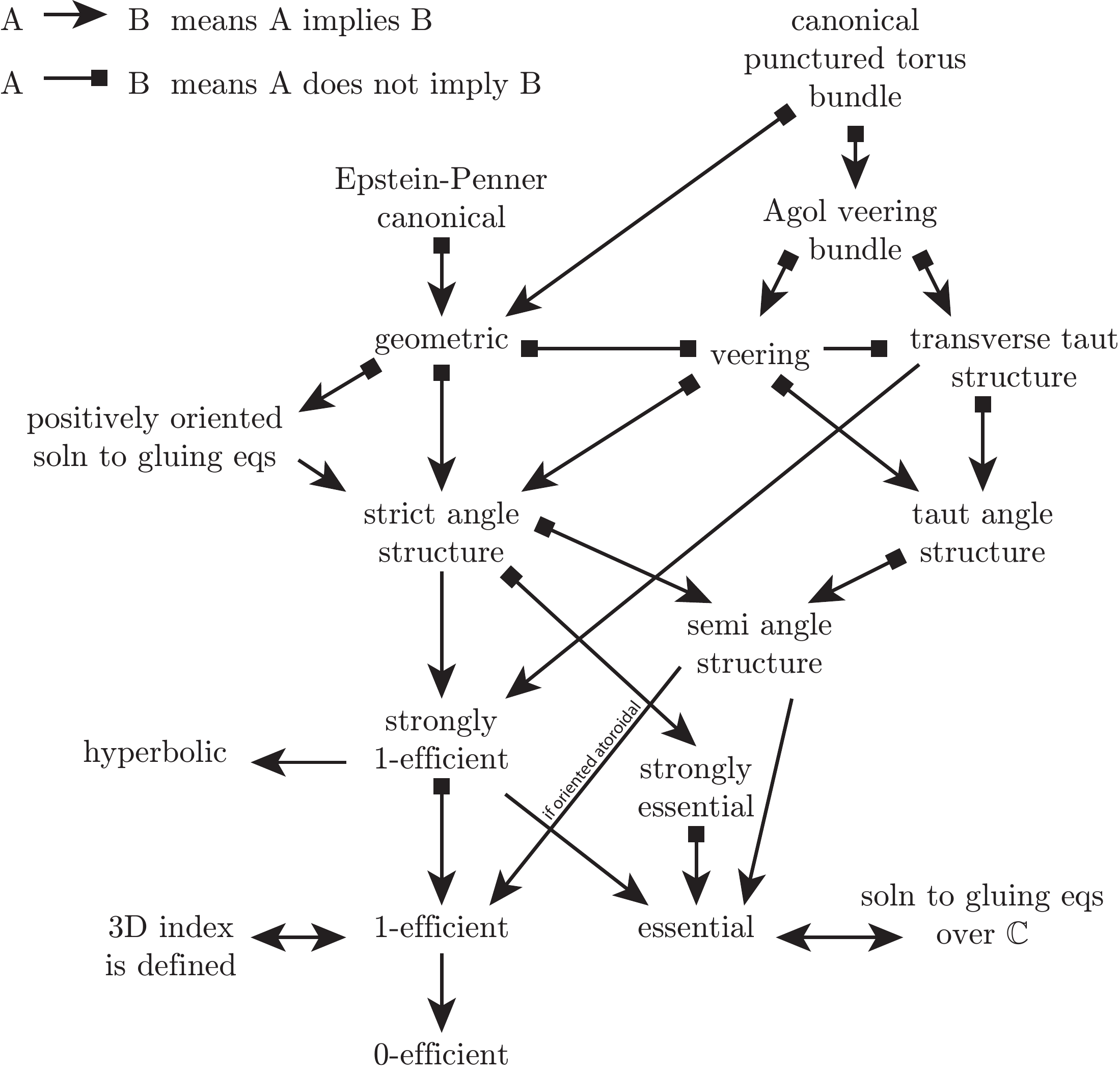}
\caption{Properties of ideal triangulations of 3-manifolds.}
\label{triangulation_properties_map}
\end{figure}

There are a number of very interesting properties that a triangulation of a manifold can have, ranging from the very strong to the very weak. Figure \ref{triangulation_properties_map} shows some of the relations between these properties, restricting now to ideal triangulations. I will survey some of these properties and their relations, and what is known about the connectivity of the subgraphs of $\mathfrak{T}(M)$ corresponding to these properties (spoiler: not much).

\begin{definition}
A \emph{geometric} triangulation has the property that the tetrahedra can be given ideal hyperbolic shapes such that they glue together to give the complete hyperbolic structure on the manifold. 
\end{definition}

The example triangulations given in Thurston's notes \cite{thurston} are geometric, and the ideal hyperbolic shapes found by solving Thurston's \emph{gluing equations}. Each tetrahedron has a shape defined by a single complex number $z\in\mathbb{C}\setminus\{0,1\}$ associated to an edge of the tetrahedron. The argument of $z$ encodes the dihedral angle of the tetrahedron at that edge, while the absolute value of $z$ determines a scaling factor. Opposite edges get the same ``complex dihedral angle'', and the other two pairs of opposite edges in a tetrahedron get complex dihedral angles $(z-1)/z$ and $1/(1-z)$. The gluing equations state that the product of the complex dihedral angles incident at an edge of the triangulation must be 1. 

\begin{definition}
A \emph{generalised angle structure} is an assignment of real numbers to the six edges of each tetrahedron of a triangulation with the following properties: 
\begin{enumerate}
\item For each tetrahedron opposite edges are assigned the same number (or \emph{angle}). 
\item The sum of the three angles in a tetrahedron is $\pi$.
\item When the tetrahedra are glued together, the sum of the angles at edges of the tetrahedra that are identified into a single edge of the triangulation is $2\pi$.
\end{enumerate}
 If we restrict the angles of a generalised angle structure to be in 
\begin{itemize}
\item $[0,\pi]$, then the generalised angle structure is a \emph{semi-angle structure}.
\item $(0,\pi)$, then the generalised angle structure is a \emph{strict angle structure}.  
\item $\{0,\pi\}$, then the generalised angle structure is a \emph{taut angle structure}.
\end{itemize}
\end{definition}
The equations satisfied by an angle structure can be thought of as a linearization of Thurston's gluing equations. 

\begin{definition}
As introduced by Jaco and Rubinstein \cite{0-efficient}, an ideal triangulation of an orientable $3$-manifold is \emph{0-efficient} if there are no embedded normal 2-spheres or one-sided projective planes. An ideal triangulation is \emph{1-efficient} if it is 0-efficient, the only embedded normal tori are vertex-linking and there are no embedded one-sided normal Klein bottles.
\end{definition}

\begin{definition}A triangulation has \emph{essential edges} if no edge loop is null-homotopic (for a one-vertex triangulation of a closed manifold), or if no ideal edge is homotopic into a vertex neighbourhood, keeping its ends in the vertex neighbourhood (for an ideal triangulation of a manifold with boundary). A triangulation has \emph{strongly essential edges} if in addition no two edges are homotopic keeping endpoints fixed (for a one-vertex triangulation of a closed manifold), or in a vertex neighbourhood (for an ideal triangulation of a manifold with boundary).
\end{definition}

\section{Connectivity}

Although Matveev/Piergallini show that $\mathfrak{T}(M)$ itself is connected, almost nothing is known about the connectivity of subgraphs of $\mathfrak{T}(M)$ corresponding to these properties. Their proofs don't seem to allow any of the control necessary to prove such connectivity. Connectivity results would be very useful, for example in constructing invariants of manifolds from functions on triangulations that are invariant under 2-3 moves. The 3D index of Dimofte, Gaiotto and Gukov  \cite{DGG2} is such a function, that is only defined on 1-efficient triangulations \cite{ghrs_index_structures}. Thus we would like to know that the subgraph of 1-efficient triangulations is connected.

A recent computer search by Craig Hodgson, Neil Hoffman, Blake Dadd and Alex Duan shows that geometric triangulations of the figure 8 knot complement are not connected, although there is an infinitely long ray in $\mathfrak{T}(M)$. So it seems that the subgraphs corresponding to the strongest properties in Figure \ref{triangulation_properties_map} are unlikely to be connected in general.

Weaker properties are more likely to correspond to connected subgraphs. As an example, consider the very weak property of not having any degree one edges. This is implied by both 1-efficiency and having essential edges. The subgraph corresponding to this property is connected. To show this, we start with a path between two triangulations without degree one edges, given to us by Matveev/Piergallini. We then modify it to avoid any introduced degree one edges. This can be achieved by a local move, inserting a particular triangulated triangular pillow in between two glued tetrahedra just before one of the three common edges is about to become degree one, then removing it immediately after the degree one edge would have ceased to exist.

This kind of trick is unlikely to work for more complicated properties. Instead, minimax techniques may work better - we would expect a minimal complexity path between two good triangulations (for some definition of complexity) to go through good triangulations.

\bibliographystyle{../../research/hamsplain}
\bibliography{../../research/henrybib}